\documentclass[10pt,conference]{IEEEtran}
\IEEEoverridecommandlockouts

\usepackage{cite}
\usepackage{amsmath,amssymb,amsfonts}
\usepackage{graphicx}
\usepackage{textcomp}
\usepackage{xcolor}
\usepackage{booktabs}
\usepackage{multirow}
\usepackage{listings}
\usepackage{colortbl}
\usepackage[caption=false,font=footnotesize]{subfig}
\usepackage{url}
\usepackage{pifont}
\usepackage{pgfplots}
\usepackage{threeparttable}
\usepackage{balance}
\usepackage{arydshln}
\usepackage{xspace}
\usepackage{tikz}
\usepackage{enumitem}
\usepackage[most]{tcolorbox}
\usepackage[ruled,vlined,linesnumbered]{algorithm2e}
\usetikzlibrary{arrows.meta,positioning,shapes.geometric,fit,calc}

\newcommand{\ourmethod}{\textsc{Trail}\xspace} 

\newcommand{\revtext}[1]{#1}

\def\BibTeX{{\rm B\kern-.05em{\sc i\kern-.025em b}\kern-.08em
    T\kern-.1667em\lower.7ex\hbox{E}\kern-.125emX}}
\begin{document}

\title{Translator vs. Challenger: Adversarial Agentic Learning for C-to-Rust Translation}

\author{
\IEEEauthorblockN{Chaofan Wang, Xiaodong Gu, Yuling Shi, Chao Hu, Beijun Shen\textsuperscript{*}\thanks{\textsuperscript{*}Beijun Shen is the corresponding author.}}
\IEEEauthorblockA{\textit{Shanghai Jiao Tong University}\\
\{chaofwang, xiaodong.gu, yuling.shi, ythere, bjshen\}@sjtu.edu.cn}
}

\maketitle

\begin{abstract}
C-to-Rust translation remains challenging due to the substantial semantic gap between the two languages. Recent experience-enhanced LLM translators improve translation quality by learning reusable insights from prior failures and repairs. Yet learned insights do not automatically constitute reusable translation knowledge: because they are derived from sparse and program-specific translation traces, they often contain missing conditions, narrow applicability boundaries, or overlooked corner cases. This limits their robustness and generalizability when applied to new translation scenarios.
We present \ourmethod, an adversarial agentic learning framework for robust C-to-Rust translation. \ourmethod employs two collaborating agents: a Translator that derives candidate insights from translation failures and accepted repairs, and a Challenger that actively searches for weaknesses, gaps, and boundary cases through adversarial challenges. 
To improve the robustness of individual insights and the completeness of insight collections, \ourmethod performs adversarial learning at two levels. Individual-insight adversarial learning repeatedly stress-tests each insight to refine its applicability conditions and constraints, while compositional insight adversarial learning strengthens groups of related insights by exposing conflicts, gaps, and uncovered corner cases. By repeatedly challenging learned insights and their compositions with executable counterexamples, \ourmethod transforms trace-specific translation experience into robust, reusable, and generalizable translation knowledge. 
We evaluate \ourmethod on two project-level benchmarks, CRUST-Bench and SmartC2Rust-Bench. Compared with the strongest LLM-based baseline, \ourmethod achieves average relative improvements of 23.1\% in syntax accuracy and 15.9\% in semantic accuracy. Furthermore, the adversarially refined insights transfer effectively across benchmarks, demonstrating strong generalizability across diverse C-to-Rust translation tasks.
\end{abstract}


\section{Introduction}

The growing adoption of Rust has created increasing demand for automated migration~\cite{c2rust-github,crust2019} of existing C codebases~\cite{rustrepotrans2024,crustbench2025}. However, C-to-Rust translation remains challenging due to the substantial semantic gap between the two languages. While C relies heavily on raw pointers, manual memory management, and implicit type conversions, Rust enforces strict ownership, borrowing, and compile-time safety guarantees. Bridging this gap requires more than syntax-level translation: translators must infer and reconstruct safety-related semantics~\cite{emre2021safer,emre2023aliasing,zhang2023ownership,hong2024dontwrite} that are often implicit in C programs. Consequently, rule-based techniques and zero-shot Large Language Model (LLM) generation frequently produce uncompilable or semantically incorrect Rust code, particularly for complex, project-specific scenarios~\cite{cai2025rustmap,evoc2rust2025,crustbench2025,smartc2rust}.
These challenges motivate translation systems that can continuously acquire and refine reusable knowledge from prior translation attempts, enabling them to handle diverse coding patterns and project-specific corner cases.

\begin{figure}[t]
    \centering
    \includegraphics[width=0.97\linewidth]{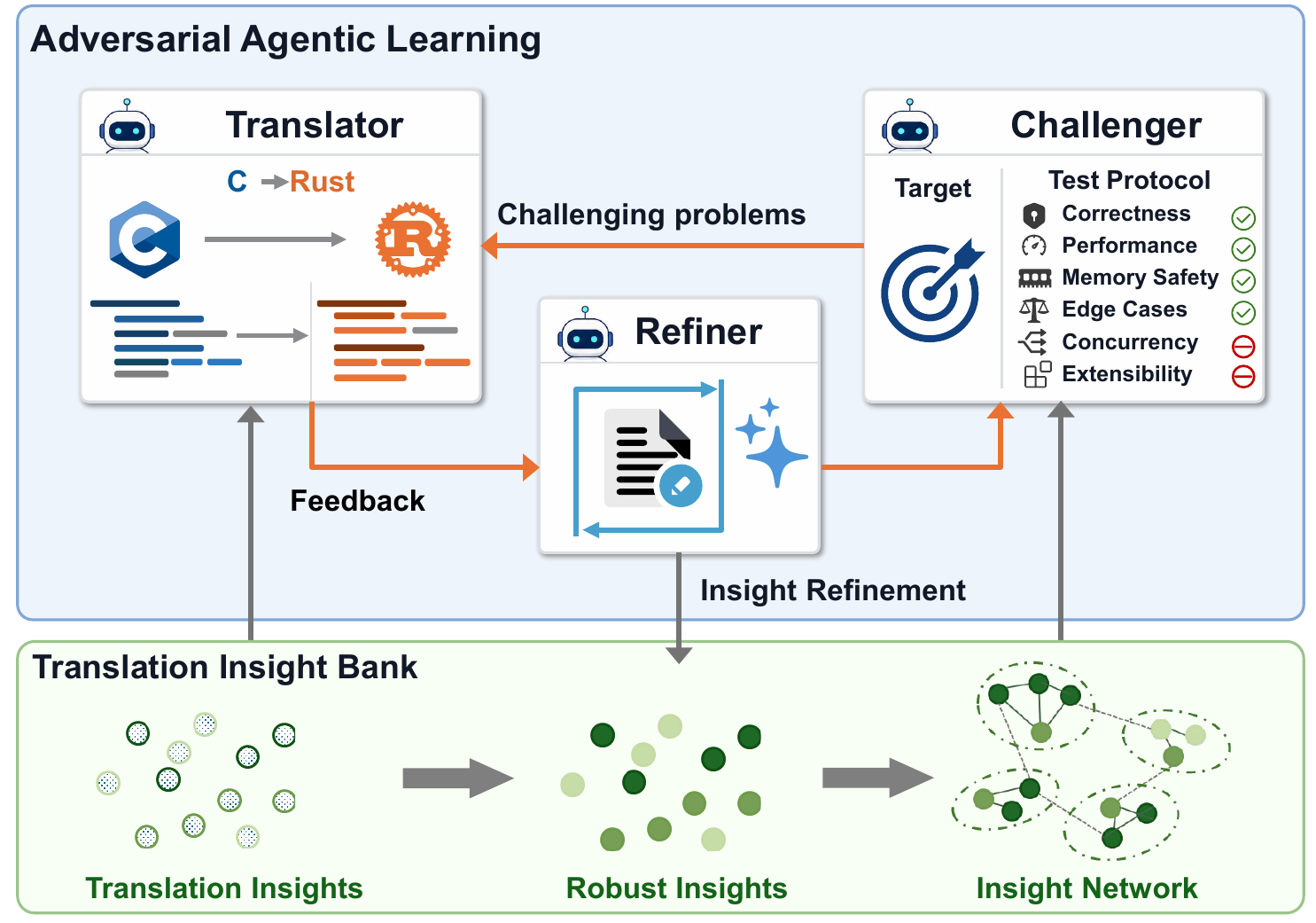}
    \caption{Illustration of adversarial agentic learning for robust and generalizable translation insights. 
    }
    \label{fig:idea}
\end{figure}

\begin{figure*}[t]
    \centering
    \includegraphics[width=0.95\linewidth, trim=0 0 0 0, clip]{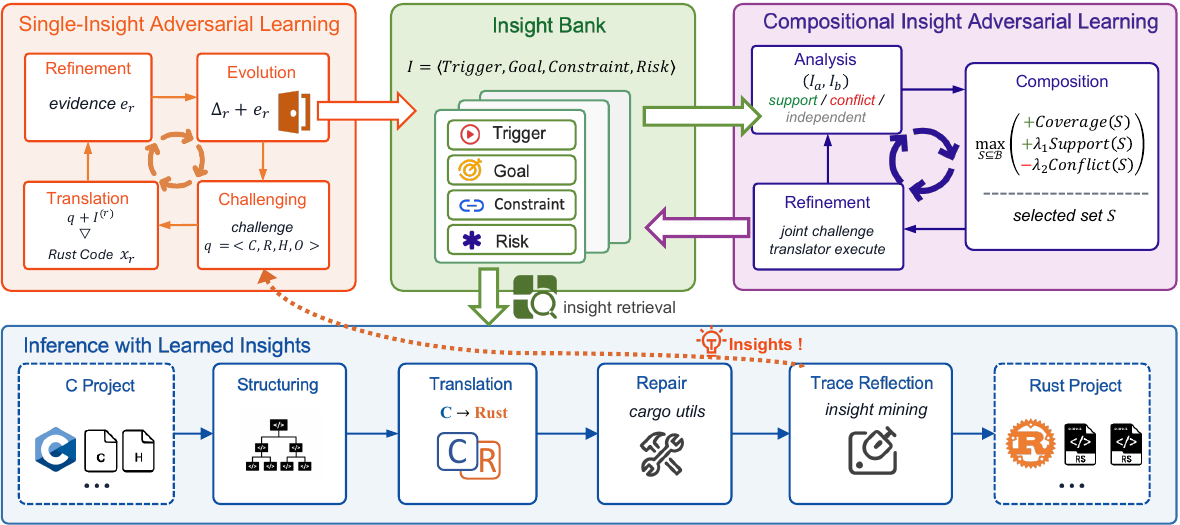}
    \caption{Overview of \ourmethod. A Challenger, Translator, and Refiner form an adversarial challenge–translate–refine loop that transforms translation experience into reusable knowledge. Adversarial refinement at both the individual-insight and compositional levels improves the robustness and completeness of learned insights for project-level C-to-Rust translation.}
    \label{fig:overview}
\end{figure*}

Existing approaches to C-to-Rust translation have increasingly shifted from manually engineered knowledge toward automatically learned translation experience. Early systems augment LLMs with handcrafted semantic mappings~\cite{evoc2rust2025} and translation guidelines~\cite{smartc2rust} to bridge the semantic gap between C and Rust. While effective for recurring patterns, such manually authored knowledge is costly to construct, difficult to maintain, and inherently limited in its coverage of real-world C codebases. Motivated by recent advances in experience-enhanced agents~\cite{reflexion2023,expel2024,agentworkflowmemory2024,sweexp2025}, newer C-to-Rust translators automatically learn reusable insights from translation failures and successful repairs~\cite{cai2025rustmap,smartc2rust,wang2026his2trans}. However, existing approaches primarily focus on accumulating learned insights, implicitly assuming that insights derived from past translation traces are reusable. In practice, because such insights originate from sparse and program-specific experiences, they often capture local heuristics rather than general translation knowledge, leaving missing conditions, narrow applicability boundaries, and overlooked corner cases. Consequently, insights that appear effective for previously observed failures may not generalize to unseen coding patterns, alternative APIs, or interactions with other translation constraints.

To address this challenge, we propose \ourmethod, an adversarial agentic learning framework that continuously refines translation insights through interaction among a Translator, a Challenger, and a Refiner. Unlike prior approaches that treat learned insights as static knowledge to be validated and stored, \ourmethod formulates insight learning as a continual adversarial process: the Translator extracts candidate insights from failures and successful repairs, while the Challenger actively constructs executable counterexamples to expose missing conditions, boundary cases, and incorrect assumptions. A Refiner then converts the resulting execution feedback into structured insight updates through iterative challenge--translate--refine cycles, progressively turning trace-specific experiences into robust, reusable knowledge. Our adversarial learning operates at two levels: at the individual level, it refines each insight to produce robust translation knowledge under diverse scenarios; at the compositional level, it jointly challenges groups of related insights to surface conflicts, gaps, and interaction-induced corner cases, organizing them into a coherent insight network that improves both coverage and consistency.
Figure~\ref{fig:idea} illustrates the overall idea of our framework.

We evaluate \ourmethod on 100 C-to-Rust translation projects using three backend LLMs. Results show that \ourmethod consistently outperforms LLM-based baselines, achieving average relative improvements of 23.1\% in syntax accuracy and 15.9\% in semantic accuracy. Furthermore, a transfer study on 20 projects from an independent benchmark shows that adversarially refined insights remain effective on unseen data, suggesting that \ourmethod learns generalizable and reusable translation knowledge.

The main contributions of this work can be summarized as:

\begin{itemize}[leftmargin=20pt, topsep=2pt]
\item We propose \ourmethod, the first adversarial agentic learning framework for C-to-Rust translation. Through adversarial interaction between a Translator and a Challenger, \ourmethod continuously refines learned insights and transforms trace-specific translation experience into reusable translation knowledge.

\item We design a two-level adversarial learning mechanism that systematically refines learned translation insights. By challenging insights both individually and compositionally, the mechanism improves their robustness, completeness, and reusability before deployment in future translations.

\item We evaluate \ourmethod on 120 C-to-Rust translation projects with multiple LLM backends. Results show that \ourmethod consistently outperforms state-of-the-art baselines and that adversarially refined insights transfer effectively across benchmarks, demonstrating strong generalizability.

\end{itemize}

\section{\ourmethod Framework}

\subsection{Framework Overview}

The core premise of \ourmethod is that C-to-Rust translation experience does not directly yield reliable reusable knowledge. Although failure-and-repair traces may expose useful heuristics, the induced insights are often partial, over-specialized to specific programs, or invalid under unseen C idioms. Instead of passively accumulating such insights in a static repository, \ourmethod formulates translation as an online adversarial learning process, in which insights are continuously mined from translation traces, challenged through executable tests, and iteratively refined during project migration.

Figure~\ref{fig:overview} illustrates the overall architecture of \ourmethod. The framework is built around structured translation insights as first-class, evolving knowledge units (Section~\ref{sec:insight}), and organized into three collaborating agents: a \emph{Challenger}, a \emph{Translator}, and a \emph{Refiner}. The Challenger synthesizes executable C-to-Rust challenges that expose missing conditions, boundary cases, or implicit assumptions in existing insights. Conditioned on selected insights, the Translator generates Rust implementations that satisfy the required interface while preserving the observable behavior of the original C program. Execution results, including compilation outcomes and test executions, provide concrete feedback, which the Refiner uses to revise the corresponding insights. Together, these agents form a closed-loop challenge--translate--refine process for continuously improving translation knowledge.

To enhance robustness and completeness, \ourmethod performs adversarial learning at two levels. At the \emph{individual-insight level}, executable challenges probe whether a single insight captures a specific C-to-Rust constraint, such as pointer–buffer coupling or ownership transfer semantics (Section~\ref{sec:single-insight}). At the \emph{compositional level}, multiple related insights are jointly evaluated under coordinated challenges to expose inconsistencies, coverage gaps, and interaction-induced corner cases (Section~\ref{sec:compositional-insight}). The resulting insight bank evolves into a structured, reusable knowledge base supporting project-level C-to-Rust translation and automated repair (Section~\ref{sec:pipeline}).

\begin{table}[t]
\centering
\caption{An Example of a Translation Insight}
\vspace{-6pt}
\small
\label{tab:insight-example}
\definecolor{InsightBack}{HTML}{F8FAFC}
\definecolor{InsightFrame}{HTML}{111827}
\definecolor{InsightTitleBack}{HTML}{050505}
\definecolor{InsightSep}{HTML}{CBD5E1}
\definecolor{InsightAccent}{HTML}{1D4ED8}
\definecolor{InsightCode}{HTML}{B42318}
\newcommand{\InsightField}[1]{\textbf{\textcolor{InsightAccent}{[#1]}}}
\newcommand{\InsightToken}[1]{\texttt{\textcolor{InsightCode}{#1}}}

\begin{tcolorbox}[
    enhanced,
    title=\textbf{Translation Insight: Pointer--Buffer Coupling},
    colback=InsightBack,
    colframe=InsightFrame,
    coltitle=white,
    colbacktitle=InsightTitleBack,
    fonttitle=\bfseries,
    fontupper=\footnotesize,
    boxrule=0.9pt,
    left=1.8mm, right=1.8mm, top=1.6mm, bottom=1.6mm,
    arc=0mm,
    titlerule=0pt,
    boxed title style={
        sharp corners,
        colframe=InsightTitleBack,
        colback=InsightTitleBack,
        left=1.8mm, right=1.8mm, top=0.8mm, bottom=0.8mm
    },
    segmentation style={
        draw=InsightSep,
        line width=0.4pt,
        dash pattern=on 3pt off 3pt
    }
]

\InsightField{Trigger}\\
C code passes raw pointers, arrays, allocated buffers, string lengths, capacities, or null terminators.


\InsightField{Goal}\\
Translate the pointer--buffer relation into Rust slices, \InsightToken{Vec}, or \InsightToken{String} while preserving length, capacity, and terminator semantics.


\InsightField{Constraint}\\
Bounds-check every index derived from C pointer arithmetic and preserve whether a terminator is included in the logical length.


\InsightField{Risk}\\
Rust code may still compile while dropping trailing-NUL handling, overrunning slices, or allocating the wrong capacity.


\InsightField{Tags}\\
\text{\scriptsize \InsightToken{pointer-buffer}, \InsightToken{ownership-transfer}, \InsightToken{nul-terminated-string}.
}

\end{tcolorbox}
\vspace{-16pt}
\end{table}

\subsection{Translation Insight Representation}
\label{sec:insight}

To enable systematic adversarial refinement, \ourmethod represents each \emph{translation insight} as a structured, executable knowledge unit that captures recurring C-to-Rust translation patterns extracted from translation traces.

Formally, an insight is defined as a 4-tuple:
\begin{equation}
I = \langle \mathit{Trigger},\; \mathit{Goal},\; \mathit{Constraint},\; \mathit{Risk} \rangle
\label{eq:insight}
\end{equation}
where \textit{Trigger} specifies the applicability conditions under which the insight is activated, \textit{Goal} characterizes the intended translation effect, \textit{Constraint} encodes semantic or structural requirements that must be preserved, and \textit{Risk} describes potential failure modes or unintended side effects induced by applying the insight. In addition, each insight is annotated with a set of lightweight semantic \texttt{tags} to facilitate retrieval and cross-insight composition. Table~\ref{tab:insight-example} presents a representative buffer-related insight. 

\subsection{Single-Insight Adversarial Learning}
\label{sec:single-insight}

\ourmethod treats each learned translation insight as a falsifiable hypothesis and subjects it to iterative adversarial testing. The objective of single-insight adversarial learning is to actively construct counterexamples that expose the boundary conditions and failure modes of an insight, thereby improving its robustness and generality.

Given an insight $I$, \ourmethod performs up to $K$ rounds of counterexample-driven refinement. In each round, the Challenger generates an executable C-to-Rust challenge targeting the applicability boundary of $I$, the Translator solves it under $I$ and the required Rust interface, and the Refiner attributes compiler or test evidence to insight deficiencies. An evidence gate then determines whether the insight is updated or retained unchanged, and the resulting version is carried to the next round. After $K$ rounds, the final refined version is admitted to the insight bank, while candidates lacking sufficient tool-observed evidence are discarded. Algorithm~\ref{alg:challenge-translate-refine} summarizes this Challenge--Translate--Refine loop.

\begin{algorithm}[t]
\small
\caption{\small Adversarial Challenge--Translate--Refine loop}
\label{alg:challenge-translate-refine}
\KwIn{Initial refinement target $Z^{(0)}$, maximum rounds $K$}
\KwOut{Refined target $Z^{\star}$}
\For{$r \leftarrow 0$ \KwTo $K-1$}{
    $q_r \leftarrow \textsc{Challenge}(Z^{(r)})$\tcp*[r]{Challenger probes a boundary}
    \If{$\neg\,\textsc{SanityCheck}(q_r, Z^{(r)})$}{
        $Z^{(r+1)} \leftarrow Z^{(r)}$\;
        \textbf{continue}\;
    }
    $x_r \leftarrow \textsc{Translate}(q_r, Z^{(r)})$\tcp*[r]{Translator uses $Z^{(r)}$ as constraint(s)}
    $e_r \leftarrow \textsc{Execute}(x_r, q_r)$\tcp*[r]{Compile and test the result}
    $\Delta_r \leftarrow \textsc{Refine}(Z^{(r)}, q_r, x_r, e_r)$\tcp*[r]{Refiner attributes failures}
    $Z^{(r+1)} \leftarrow \textsc{EvidenceGate}(Z^{(r)}, \Delta_r, e_r)$\tcp*[r]{Accept supported updates only}
}
$Z^{\star} \leftarrow Z^{(K)}$\;
\Return{$Z^{\star}$}\;
\end{algorithm}

Specifically, each adversarial refinement round proceeds through four steps:

\subsubsection{\textbf{Challenge Generation}} 
The goal of challenge generation is to construct test cases that maximize the likelihood of falsifying an incomplete insight. 
Given the current insight instance $I^{(r)}=\langle T_I,G_I,\Phi_I,\rho_I\rangle$, the Challenger first derives an activation predicate $a_I=\textsc{Predicate}(T_I)$ that identifies programs capable of triggering the targeted transformation behavior, and extracts adversarial boundary conditions $b_I=\textsc{Boundary}(\Phi_I,\rho_I)$ that characterize critical edge cases implied by the insight specification. 

Based on $a_I$ and $b_I$, the Challenger instantiates a C-to-Rust challenge $q=\langle C_q,R_q,H_q,O_q\rangle$, where $C_q$ specifies the C program exercising the targeted idiom, $R_q$ fixes the Rust-facing interface constraints, $H_q$ defines the execution harness for invoking the translated implementation, and $O_q$ encodes the expected C semantics as executable assertions. 
By combining activation conditions with boundary constraints under both source-language semantics and target-interface requirements, the generated challenge is designed to expose missing preconditions, over-generalized assumptions, and unhandled corner cases in the current insight.

Before execution, \ourmethod applies three sanity checks to filter generated challenges, ensuring execution budget is spent on valid and diagnostic cases. \textsc{Leak} removes duplicates and near-duplicates, \textsc{Activates} verifies triggers and boundary conditions, and \textsc{ToolCheck} ensures executability by checking that the C snippet runs, the Rust harness type-checks against a stub for $R_q$, and the oracle passes under \texttt{cargo test}.

\subsubsection{\textbf{Insight-Constrained Translation}}
Each challenge is evaluated by the Translator with the current insight $I^{(r)}$ injected as an active constraint. Conditioned on $I^{(r)}$, it synthesizes Rust code that satisfies the required interface $R_q$ while preserving the observable behavior of $C_q$.

The resulting implementation is validated via \texttt{cargo check} and \texttt{cargo test}. If both succeed, the challenge is marked as solved, and \ourmethod records the implementation and passing trace as positive evidence that $I^{(r)}$ covers the targeted behavior. Otherwise, compilation or test failures are treated as counterexamples; the failed program, compiler diagnostics, interface mismatches, and assertion failures are forwarded to the Refiner to update $I^{(r)}$.

\subsubsection{\textbf{Failure-Driven Refinement}} 
When a challenge reveals a failure, the Refiner performs failure attribution to establish a diagnostic link between observed errors and the targeted insight. Insight updates are proposed only when grounded in concrete evidence, including compiler diagnostics, failing assertions, or execution traces.

Attributed failures are categorized into four insight-level deficiencies, each mapped to a field-level update. A \emph{Trigger gap} denotes missing applicability conditions (e.g., requiring a visible length or terminator for a pointer rule), refining the \emph{Trigger}. A \emph{Goal mismatch} indicates an incorrect or incomplete objective (e.g., preserving bytes but violating the Rust-facing API contract), revising the \emph{Goal}. A \emph{Constraint violation} captures insufficient or overly permissive constraints (e.g., unchecked slice construction or violated ownership assumptions), strengthening the \emph{Constraint}. A \emph{Missing risk} records unmodeled failure modes (e.g., passing \texttt{cargo check} while mishandling trailing NUL bytes, introducing unintended aliasing, or altering allocation capacity), extending the \emph{Risk} with explicit guards or failure conditions.

Refinement is triggered only by actionable failures; translator noise, invalid challenges, and successful executions do not produce updates.

\subsubsection{\textbf{Evidence-Gated Insight Evolution}} 
The evidence gate validates each refinement proposal via replay-based verification. Given a proposed update $\Delta_r$, \ourmethod applies it to the current insight $I^{(r)}$ to obtain a candidate $\hat{I}^{(r+1)}$. The Translator then re-executes the same challenge under $\hat{I}^{(r+1)}$ as an active constraint.

If the replay succeeds, the update is accepted and $\hat{I}^{(r+1)}$ becomes the insight for the next round. Otherwise, the proposal is rejected and $I^{(r)}$ is retained. Insights remain unchanged when (i) the original challenge already succeeds under $I^{(r)}$, or (ii) no refinement proposal is generated by the Refiner. After the final round, the latest accepted version is taken as the refined insight.

\subsection{Compositional Insight Adversarial Learning}
\label{sec:compositional-insight}

Improving individual insights does not guarantee that they form a complete and coherent body of translation knowledge. In project-level C-to-Rust translation, multiple insights are often activated simultaneously along shared call chains or data flows. Although each insight may be correct in isolation, their composition can introduce conflicts, missing coordination constraints, or corner cases that are not observable in single-insight refinement. For example, an insight preserving a C terminator may conflict with another constructing a Rust slice excluding it; similarly, ownership-transfer rules may interfere with insights treating the same buffer as a borrowed view. Thus, locally correct insights may lead to inconsistent global behavior.

To address this limitation, \ourmethod performs adversarial learning at the compositional level. It jointly challenges co-activated insight sets within a module and evaluates their combined constraints. Through joint conflict detection and coverage-aware set refinement, \ourmethod determines safe composability, required precedence or applicability constraints, and interaction-induced translation principles.

Specifically, compositional adversarial learning proceeds through three steps:

\subsubsection{\textbf{Counterfactual Interaction Analysis}}
Starting from the insight bank $\mathcal{B}$, whose entries have passed single-insight adversarial learning, \ourmethod analyzes pairwise interactions among insights that may co-activate in the same C-to-Rust context. For an insight pair $(I_a, I_b)$, \ourmethod performs counterfactual reasoning~\cite{zeng2025pruning,DBLP:conf/acl/ZengZSHCSG26,zhang2026paratempo} over their \emph{Trigger}, \emph{Goal}, \emph{Constraint}, \emph{Risk}, C-idiom tags, and Rust-facing obligations to classify their relation as \emph{support}, \emph{independent}, or \emph{conflict}. The analysis asks whether removing, weakening, or reordering one insight would alter the Rust interface, ownership model, aliasing assumptions, or executable behavior required by the other, yielding support and conflict scores $s_{ab}, c_{ab} \in [0,1]$ together with explanations.

Predicted conflicts are not immediately rejected. Instead, they are flagged for adversarial testing to distinguish genuine inconsistencies from cases that require refined applicability conditions, strengthened constraints, or explicit precedence rules. This is necessary because apparent conflicts often arise from conflating internal C representation requirements (e.g., preserving a terminator in an allocated buffer) with external Rust API obligations (e.g., exposing only initialized data via a slice).

\subsubsection{\textbf{Coverage-Aware Insight Composition}}
Given the interaction graph, \ourmethod constructs compact insight sets for joint adversarial testing. Each set is expected to represent a coherent C-to-Rust feature, and co-activate during module-level translation. Candidate sets are formed from related triggers and overlapping language-idiom tags, such as \texttt{pointer-buffer}, \texttt{ownership-transfer}, \texttt{nul-terminated-string}, and \texttt{api-preservation}.

To balance coverage and compatibility, \ourmethod selects an insight set $S \subseteq \mathcal{B}$ by maximizing

\begin{equation}
\scalebox{0.9}{$\displaystyle
\max_{S \subseteq \mathcal{B}} \; \underbrace{\mathit{Coverage}(S)}_{\text{C-to-Rust tag coverage}} + \lambda_1 \underbrace{\mathit{Support}(S)}_{\text{LLM support score}} - \lambda_2 \underbrace{\mathit{Conflict}(S)}_{\text{LLM conflict score}}
$}
\label{eq:selection}
\end{equation}
where $\mathit{Coverage}(S)$ measures normalized coverage of distinct C-to-Rust semantic tags, defined as $|\cup_{I\in S}\mathit{tags}(I)|/\min(B,|\mathcal{T}|)$. Here $B$ denotes the maximum number of insights per joint challenge, and $\mathcal{T}$ is the tag vocabulary of the candidate set. For $|S|>1$, $\mathit{Support}(S)$ and $\mathit{Conflict}(S)$ are defined as average pairwise LLM scores:
\begin{equation}
\begin{aligned}
\mathit{Support}(S)&=\frac{2}{|S|(|S|-1)}\sum_{a<b}s_{ab},\\
\mathit{Conflict}(S)&=\frac{2}{|S|(|S|-1)}\sum_{a<b}c_{ab}.
\end{aligned}
\end{equation}
For singleton sets, both terms are zero.

In practice, \ourmethod constructs compositions via greedy optimization of Eq.~\ref{eq:selection}. This approximation suffices as the objective is interpretable adversarial evaluation rather than global subset optimality. The resulting set captures realistic translation scenarios, such as helper chains that compute buffer lengths, enforce terminators, transfer ownership, and preserve stable Rust-facing APIs.

\subsubsection{\textbf{Compositional Adversarial Refinement}}
The selected composition $S$ is refined via joint adversarial testing. The Challenger generates executable C-to-Rust challenges that jointly activate all insights in $S$ and force oracle dependence on their interaction; sanity checks reject cases where insights are exercised independently. Valid challenges combine multiple interacting C-to-Rust concerns, requiring a unified Rust solution satisfying all insights.

The Translator attempts the challenge with full activation of $S$, and execution produces compiler and test evidence as in the single-insight setting. Success indicates safe co-activation in shared contexts. Failures are attributed by the Refiner to conflicting constraints, missing coordination conditions, incomplete precedence relations, or interaction-specific corner cases. The evidence gate admits only verdict-supported updates, including field refinements, relation updates, co-activation preconditions, or precedence rules. For instance, ownership transfer may override borrowed-view assumptions when the C caller releases the buffer, while slice construction excludes a terminator preserved internally. If a failure exposes a non-representable pattern, the Refiner introduces a new candidate insight, which is re-entered into single-insight adversarial learning before entering the bank.

Through iterative joint challenges and evidence-driven refinement, \ourmethod improves interaction coherence and the completeness of learned C-to-Rust translation knowledge.

\subsection{Inference with Learned Insights}
\label{sec:pipeline}

Building on the adversarially refined insight bank, \ourmethod performs project-level C-to-Rust translation as a closed-loop inference process that interleaves insight utilization and online continual learning. The inference pipeline is implemented by three cooperating agents: the Translator, which generates initial Rust implementations; the Repairer, which fixes \texttt{cargo check} and \texttt{cargo test} failures; and the Reflector, which distills reusable translation insights from execution traces and repair trajectories.

Given a C project, \ourmethod first applies macro expansion and constructs a dependency graph to derive a topological order over functions. For each function, context-relevant insights are retrieved to guide both translation and subsequent repair. The generated Rust code is first validated and iteratively corrected using compilation feedback, then integrated into the project and further improved based on test execution signals.
To close the learning loop, the Reflector mines reusable insights from translation and repair traces, continuously enriching the insight bank with newly observed patterns and failure-driven refinements.

Specifically, the pipeline consists of five stages.

\subsubsection{\textbf{Project Structuring for Translation}}
\ourmethod first constructs a translation-ready project representation and establishes a dependency-aware translation order. It macro-expands the source code and performs whole-project analysis to identify translation targets, shared declarations, Rust interfaces, and inter-function dependencies. The resulting dependency graph is topologically sorted to determine the subsequent function translation order.

\subsubsection{\textbf{Insight Retrieval}}
Before translating each function, \ourmethod retrieves context-relevant insights from the insight bank using the target C function, Rust interface, macro-expanded surrounding context, available Rust artifacts, and prior translation or repair feedback as the query.

Retrieval proceeds in two steps. Candidate insights are first selected by matching their \emph{Trigger} and \texttt{tags} against the query, and then re-ranked by contextual relevance, compatibility, and conflict risk estimated from compositional adversarial learning. The top-$N$ compatible insights are retained, while high-conflict combinations are filtered out.

\subsubsection{\textbf{Knowledge-Guided Translation}}
For each function, \ourmethod combines dependency-aware project context with retrieved insights. The context includes the target Rust interface, aligned macro-expanded C code, relevant declarations and constants, validated Rust artifacts from previously translated functions, and essential dependency information, while unrelated files are omitted or abstracted to meet prompt constraints.

Conditioned on this context, the Translator generates Rust code that conforms to the target interface and preserves the observable behavior of the original C function. Retrieved insights provide structured guidance on semantic constraints, translation patterns, and potential risks.

\subsubsection{\textbf{Execution-Guided Repair}}
After translation, \texttt{cargo check} is used for validation. If it succeeds, the implementation is accepted and the pipeline proceeds; otherwise, the Repairer is invoked. It performs localized fixes using compiler diagnostics, the target Rust interface, relevant C context, previous Rust attempts, and retrieved insights. This repair loop repeats until compilation succeeds or the budget is exhausted.

Once all functions compile, the integrated project is validated using \texttt{cargo test}. If tests pass, the translation is accepted; otherwise, the Repairer localizes failures using test outputs, panic traces, assertions, dependency context, and recently translated components, and performs further targeted repairs under the same constraints.

\subsubsection{\textbf{Trace Reflection for Insight Mining}}
\ourmethod performs reflection over all execution traces. The Reflector first filters out traces that are project-specific, non-recurring, or low-signal traces. For the remaining traces, it analyzes C context, compiler diagnostics, test failures, generated candidates, and final outcomes to extract underlying conditions, objectives, constraints, or risks, which are summarized as candidate insights in the form $\langle \mathit{Trigger}, \mathit{Goal}, \mathit{Constraint}, \mathit{Risk} \rangle$.

These candidates are then screened to remove duplicates, triggerless patterns, contradictions with existing insights, or artifacts tied to project-specific structure. The remaining candidates are forwarded for adversarial learning before entering the insight bank, completing the translation–learning loop.

\section{Experimental Setup}


We conduct experiments to evaluate the effectiveness of \ourmethod, aiming to answer the following research questions:

\begin{itemize}
\item \textbf{RQ1 (Overall Effectiveness):} How does \ourmethod perform compared with state-of-the-art baselines on project-level C-to-Rust translation?

\item \textbf{RQ2 (Ablation of Key Components):} What is the contribution of insight mining and the two-level adversarial learning mechanism to translation performance?

\item \textbf{RQ3 (Generalization of Learned Insights):} How well do insights learned from one set of C-to-Rust projects transfer to unseen project collections in improving translation performance?

\item \textbf{RQ4 (Sensitivity Analysis):} How do key hyperparameters, including insight selection size and the number of adversarial learning rounds, affect the performance and stability of \ourmethod?
\end{itemize}

\subsection{Datasets}

We evaluate \ourmethod on two C-to-Rust translation benchmarks.

\subsubsection{CRUST-Bench}
CRUST-Bench~\cite{crustbench2025} contains 100 real-world C projects collected from GitHub, Linux, and PostgreSQL, covering domains such as data structures, cryptography, encoding, parsing, and system utilities. Each project is paired with an idiomatic Rust interface crate and executable tests. We use CRUST-Bench as the primary benchmark for effectiveness, ablation, and sensitivity studies.

\subsubsection{SmartC2Rust-Bench}
To evaluate cross-benchmark generalization, we use the benchmark introduced by SmartC2Rust~\cite{smartc2rust}. Of its 21 programs, 20 are publicly available and used in our experiments. These subjects comprise small-to-medium command-line utilities and library-style programs with build scripts, test scripts, and entry-point specifications. Unlike CRUST-Bench, which relies on Rust interface crates, SmartC2Rust-Bench evaluates translations through executable behavioral checks.

\subsection{Models}

We evaluate \ourmethod with three representative LLM backends: \textbf{GPT-5.4-mini}, a compact model with strong code-generation capabilities; \textbf{Kimi-K2.5}, a large-scale model with an extended context window; and \textbf{DeepSeek-V4-Flash}, a fast inference model optimized for coding tasks. 
All models are accessed through OpenAI-compatible APIs. Unless otherwise specified, we use \texttt{temperature\,=\,1.0} and \texttt{max\_output\_tokens\,=\,128000} for all experiments.

\subsection{Baselines}

We compare \ourmethod against five representative baselines spanning rule-based, hybrid, and LLM-based C-to-Rust translation systems. All baselines are evaluated using their default configurations.

\begin{itemize}
\item \textbf{C2Rust}~\cite{c2rust-github} is a rule-based source-to-source translator that mechanically converts C code into Rust. It serves as a non-LLM baseline for transpilation-based migration.

\item \textbf{C2SaferRust}~\cite{alqadiri2025translating} is a hybrid approach that first translates C into unsafe Rust using C2Rust and then employs an LLM to incrementally rewrite unsafe code into safer Rust while preserving behavior through test-based verification.

\item \textbf{Direct Prompting} uses the backend LLM to translate each callable without learned insights or iterative repair, representing the model's zero-experience translation capability.

\item \textbf{Self-Repair}~\cite{crustbench2025} augments direct translation with iterative compilation-driven repair (up to five rounds) but does not leverage learned insights, isolating the effect of repair-based feedback.

\item \textbf{SmartC2Rust}~\cite{smartc2rust} is an LLM-based translation framework that performs translation and repair using compilation and test feedback, but does not incorporate structured translation insights.

\end{itemize}

No public C-to-Rust baseline provides a reproducible structured-insight pipeline for comparison. Therefore, in RQ2 we include a variant with directly mined insights via trace reflection, serving as a non-adversarial baseline. 

\begin{table*}[t]
\centering
\caption{Overall Comparison of \ourmethod and Baselines on 100 CRUST-Bench Projects}
\label{tab:main-results}
\small
\renewcommand{\arraystretch}{1.02}
\begin{tabular}{@{}l@{\hspace{1.35em}}lccccc@{}}
\toprule
\multirow{2}{*}{\textbf{Model}} & \multirow{2}{*}{\textbf{Method}} & \multirow{2}{*}{\textbf{CompRate}} & \multirow{2}{*}{\textbf{TestRate}} & \multirow{2}{*}{\textbf{SafeRate}} & \multicolumn{2}{c}{\textbf{Idiomaticity}} \\
\cmidrule(lr){6-7}
 & & & & & \textbf{Lint Pass $\uparrow$} & \textbf{Idiom Penalty $\downarrow$} \\
\midrule
\multicolumn{7}{@{}l}{\cellcolor{gray!15}\textit{Rule-based and hybrid baselines}} \\
\addlinespace[2pt]
\textemdash & C2Rust & 98\% & 98\% & 0.13\% & 98 & 49.06 \\
gpt-5.4-mini & C2SaferRust & 98\% & 98\% & 12.28\% & 94 & 49.20 \\
\midrule
\multicolumn{7}{@{}l}{\cellcolor{gray!15}\textit{LLM-based methods}} \\
\addlinespace[2pt]
\multirow{4}{*}{gpt-5.4-mini} & Direct Prompting & 32\% & 24\% & 99.66\% & 34 & 11.98 \\
 & Self-Repair & 56\% & 55\% & \textbf{99.67\%} & 65 & 8.70 \\
 & SmartC2Rust & 66\% & 40\% & 99.41\% & 67 & 4.77 \\
 & \textbf{\ourmethod (ours)} & \textbf{88\%} & \textbf{69\%} & 96.63\% & \textbf{88} & \textbf{3.95} \\
\midrule
\multirow{4}{*}{kimi-k2.5} & Direct Prompting & 39\% & 25\% & 98.95\% & 41 & 13.27 \\
 & Self-Repair & 76\% & 61\% & \textbf{99.03\%} & 84 & 8.48 \\
 & SmartC2Rust & 59\% & 37\% & 97.16\% & 60 & 5.21 \\
 & \textbf{\ourmethod (ours)} & \textbf{79\%} & \textbf{67\%} & 95.12\% & \textbf{89} & \textbf{4.47} \\
\midrule
\multirow{4}{*}{deepseek-v4-flash} & Direct Prompting & 21\% & 14\% & 98.81\% & 25 & 11.89 \\
 & Self-Repair & 56\% & 48\% & \textbf{99.08\%} & 67 & 9.23 \\
 & SmartC2Rust & 31\% & 26\% & 98.49\% & 15 & 7.22 \\
 & \textbf{\ourmethod (ours)} & \textbf{74\%} & \textbf{54\%} & 97.57\% & \textbf{88} & \textbf{4.22} \\
\bottomrule

\end{tabular}

\vspace{2pt}
\parbox{0.8\linewidth}{\footnotesize
$^{*}$ For LLM-based methods, results are grouped by backbone model, with best in \textbf{bold}.}
\vspace{-3pt}
\end{table*}

\subsection{Metrics}

We assess translation quality from four perspectives: compilation success, behavioral correctness, safety, and idiomaticity.

1) \textbf{CompRate}: The percentage of projects that successfully pass \texttt{cargo check}, indicating successful project-level compilation.

2) \textbf{TestRate}: The percentage of projects that pass all \texttt{cargo test} cases, indicating preservation of observable program behavior.

3) \textbf{SafeRate}: The percentage of translated callables that contain no \texttt{unsafe} blocks, measuring the extent to which translations leverage Rust's safety guarantees.

4) \textbf{Idiomaticity}: Measured using two metrics. \textbf{Lint Pass} counts the number of projects that pass \texttt{cargo clippy}, reflecting adherence to Rust coding conventions. \textbf{Idiom Penalty} follows the non-idiomatic pattern penalty defined by Rustine~\cite{rustine2025}. Higher Lint Pass and lower Idiom Penalty indicate more idiomatic Rust code.

\subsection{Implementation Details}

The insight bank is stored as structured JSON records and indexed using \texttt{FAISS}. Each insight’s \emph{Trigger} and \texttt{tags} are encoded with \texttt{BAAI/bge-base-en-v1.5} for similarity-based retrieval, with a default retrieval budget of $N=5$ insights.

All agents are implemented as structured LLM calls with fixed schemas, prompts, output formats, and acceptance criteria, ensuring consistency across the full \ourmethod pipeline and all ablation variants. All prompts are released in our artifact repository to support reproducibility.

For compositional adversarial learning, the backend LLM estimates support and conflict scores in Eq.~\ref{eq:selection}. We set $\lambda_1=\lambda_2=1$ and use a conflict threshold $\tau_c=0.7$. Candidate insights are grouped by related triggers and overlapping \texttt{tags}, and compositions are constructed greedily by iteratively adding the largest positive marginal gain until reaching $B=3$, exhausting positive-gain candidates, or encountering only unresolved conflicts.

Unless otherwise specified, each accepted insight undergoes at most $K=5$ adversarial refinement rounds. During execution-guided repair, compilation-level fixes are limited to five attempts per callable (following Self-Repair), and project-level test repair is capped at three attempts. A repair is accepted only if it succeeds or reduces the number of diagnostics or failing tests.

\section{Results}
\subsection{RQ1: Overall Effectiveness}
\label{sec:rq1}

We evaluate \ourmethod against rule-based, LLM-based, and hybrid baselines on 100 CRUST-Bench projects. Table~\ref{tab:main-results} summarizes the results.

Across all three backbone models, \ourmethod achieves the strongest LLM-based project-level correctness. With gpt-5.4-mini, it reaches 88\% CompRate and 69\% TestRate, surpassing SmartC2Rust's 66\% CompRate and Self-Repair's 55\% TestRate by 22 and 14 percentage points, respectively. Similar gains appear on kimi-k2.5 (79\%/67\% vs. 76\%/61\%) and deepseek-v4-flash (74\%/54\% vs. 56\%/48\%), showing that the effect is consistent across backbones.

\begin{table*}[t]
\centering
\caption{Ablation Study of the Proposed Insight Learning Mechanism, with Cumulative Removal of Each Component from \ourmethod}
\vspace{-3pt}
\label{tab:ablation}
\setlength{\tabcolsep}{3pt}
\renewcommand{\arraystretch}{1.05}
\resizebox{\textwidth}{!}{%
\begin{tabular}{llccccc}
\toprule
\multirow{2}{*}{\textbf{Model}} & \multirow{2}{*}{\textbf{Variant}} & \multirow{2}{*}{\textbf{CompRate}} & \multirow{2}{*}{\textbf{TestRate}} & \multirow{2}{*}{\textbf{SafeRate}} & \multicolumn{2}{c}{\textbf{Idiomaticity}} \\
\cmidrule(lr){6-7}
 & & & & & \textbf{Lint Pass $\uparrow$} & \textbf{Idiom Penalty $\downarrow$} \\
\midrule
\multirow{4}{*}{gpt-5.4-mini}
  & \textbf{\ourmethod} & \textbf{88\%} & \textbf{69\%} & \textbf{96.63\%} & 88 & \textbf{3.95} \\
  & \qquad $w/o$ Compositional Insight Adversarial Learning & 86\% & 66\% & 96.22\% & 89 & 4.00 \\
  & \qquad $w/o$ Single Insight Adversarial Learning & 85\% & 63\% & 96.60\% & \textbf{91} & 4.07 \\
  & \qquad $w/o$ Insight Mining & 75\% & 56\% & 93.87\% & 87 & 4.16 \\
\midrule
\multirow{4}{*}{kimi-k2.5}
  & \textbf{\ourmethod} & \textbf{79\%} & \textbf{67\%} & \textbf{95.12\%} & \textbf{89} & 4.47 \\
  & \qquad $w/o$ Compositional Insight Adversarial Learning & 76\% & 63\% & 94.76\% & 86 & 4.37 \\
  & \qquad $w/o$ Single Insight Adversarial Learning & 69\% & 54\% & 94.95\% & \textbf{89} & \textbf{4.28} \\
  & \qquad $w/o$ Insight Mining & 70\% & 49\% & 94.98\% & 87 & 4.31 \\
\midrule
\multirow{4}{*}{deepseek-v4-flash}
  & \textbf{\ourmethod} & 74\% & \textbf{54\%} & 97.57\% & 88 & 4.22 \\
  & \qquad $w/o$ Compositional Insight Adversarial Learning & \textbf{75\%} & 51\% & \textbf{97.74\%} & \textbf{90} & 4.26 \\
  & \qquad $w/o$ Single Insight Adversarial Learning & 70\% & 49\% & 97.73\% & 87 & 4.22 \\
  & \qquad $w/o$ Insight Mining  & 68\% & 45\% & 96.85\% & 88 & \textbf{4.10} \\
\bottomrule
\end{tabular}
}

\vspace{2pt}
\parbox{0.98\linewidth}{\footnotesize
$^{*}$ Best results per model are in \textbf{bold}.}
\vspace{-3pt}
\end{table*}

The improvements over SmartC2Rust and Self-Repair suggest that proactive knowledge injection complements iterative repair. \ourmethod mines and refines reusable insights from traces, helping prevent recurring failures rather than only fixing them after they appear.

Rule-based and hybrid baselines show a different trade-off. C2Rust and C2SaferRust reach up to 98\% CompRate and TestRate, but rely on unsafe-first transpilation and therefore suffer extremely low safety and poor idiomaticity. Thus, compile/test success alone does not characterize translation quality. Although \ourmethod does not yet match their raw pass rates, it narrows the gap while avoiding their severe safety and idiomaticity degradation.

\begin{tcolorbox}[enhanced, width=\linewidth, boxrule=0.8pt, left=2pt, right=2pt, top=2pt, bottom=2pt, drop fuzzy shadow=black,]
\textbf{Answer to RQ1.} \revtext{\ourmethod achieves the best overall effectiveness among LLM-based approaches across all three backend models, while preserving a better balance across functional correctness, safety, and idiomaticity than the unsafe transpilation-first baselines.}
\end{tcolorbox}

\subsection{RQ2: Ablation of Key Components}
\label{sec:rq2}

To quantify each component’s contribution, we conduct a stepwise ablation study on CRUST-Bench. Starting from the full \ourmethod framework, we progressively remove compositional insight adversarial learning, single-insight adversarial learning, and insight mining. Table~\ref{tab:ablation} summarizes the results.

All components are beneficial, with performance degrading under each removal. Removing compositional insight adversarial learning reduces behavioral correctness across backbones, highlighting its role in coordinating related insights. In the full system, it boosts TestRate to 69\%, 67\%, and 54\% on gpt-5.4-mini, kimi-k2.5, and deepseek-v4-flash while keeping CompRate stable. Without it, cross-insight inconsistency leads to TestRate drops and minor CompRate trade-offs (e.g., 75\%$\rightarrow$74\% on deepseek-v4-flash).

Further removing single-insight adversarial learning causes larger degradation: on kimi-k2.5, TestRate/CompRate decrease from 63\%/76\% to 54\%/69\%; on deepseek-v4-flash, from 51\%/75\% to 49\%/70\%; gpt-5.4-mini shows a smaller but consistent decline (66\%/86\%$\rightarrow$63\%/85\%), indicating that counterexample-driven refinement improves per-insight precision and robustness.

Removing insight mining yields the largest drop, with TestRate falling to 56\%, 49\%, and 45\% on gpt-5.4-mini, kimi-k2.5, and deepseek-v4-flash. This suggests that even unrefined trace-reflected insights provide useful semantic constraints beyond a standard translation-and-repair pipeline.

Across variants, SafeRate remains stable (96.63\%, 95.12\%, 97.57\%), and idiomaticity changes marginally, indicating functional gains do not compromise safety or code quality.

\begin{tcolorbox}[enhanced, width=\linewidth, boxrule=0.8pt, left=2pt, right=2pt, top=2pt, bottom=2pt, drop fuzzy shadow=black,]
\textbf{Answer to RQ2.} \revtext{Failure-driven mined insights already provide measurable benefits, while adversarial learning at both the single-insight and compositional levels further improves the reliability of reusable constraints, mainly by boosting TestRate without sacrificing SafeRate or idiomaticity.}
\end{tcolorbox}

\subsection{RQ3: Generalization of Learned Insights}
\label{sec:rq3}

To evaluate cross-benchmark generalization, we transfer the CRUST-Bench insight bank to SmartC2Rust-Bench. Table~\ref{tab:transfer-results} compares \textit{Base} without insight injection, \textit{Direct} with target-benchmark insights, and \textit{Transferred} with CRUST-Bench insights and no further adaptation.

\textit{Transferred} improves over \textit{Base} across models. For gpt-5.4-mini, CompRate/TestRate rises from 80\%/55\% to 85\%/70\%; for kimi-k2.5, CompRate stays at 90\% while TestRate improves from 65\% to 70\%; for deepseek-v4-flash, CompRate rises from 75\% to 80\% while TestRate remains 70\%. These gains indicate that the learned insights capture reusable C-to-Rust constraints beyond the source benchmark.

However, \textit{Transferred} remains below \textit{Direct}, which learns on SmartC2Rust-Bench itself. Compared with \textit{Transferred}, \textit{Direct} improves CompRate by 5 points for all models and boosts TestRate by 5, 10, and 10 points on gpt-5.4-mini, kimi-k2.5, and deepseek-v4-flash. Thus, transferred insights provide strong priors, but adversarial refinement is still needed for benchmark-specific APIs, failure modes, and boundary conditions.

\begin{table}[t]
\centering
\caption{Cross-benchmark Generalization of Adversarially Refined Insights}
\label{tab:transfer-results}
\setlength{\tabcolsep}{2.5pt}
\renewcommand{\arraystretch}{1.08}
\resizebox{\columnwidth}{!}{%
\begin{tabular}{@{}llccccc@{}}
\toprule
\multirow{2}{*}{\textbf{Model}} & \multirow{2}{*}{\textbf{Setting}} & \multirow{2}{*}{\textbf{CompRate}} & \multirow{2}{*}{\textbf{TestRate}} & \multirow{2}{*}{\textbf{SafeRate}} & \multicolumn{2}{c}{\textbf{Idiomaticity}} \\
\cmidrule(l){6-7}
 & & & & & \textbf{Lint Pass $\uparrow$} & \textbf{Idiom Penalty $\downarrow$} \\
\midrule
\multirow{3}{*}{gpt-5.4-mini}
 & Base & 80\% & 55\% & 98.57\% & 13/20 & 5.17 \\
 & Direct & \textbf{90\%} & \textbf{75\%} & 98.44\% & \textbf{16/20} & \textbf{4.72} \\
 & Transferred & 85\% & 70\% & \textbf{98.62\%} & 13/20 & 5.28 \\
\midrule
\multirow{3}{*}{kimi-k2.5}
 & Base & 90\% & 65\% & 97.62\% & 14/20 & 19.13 \\
 & Direct & \textbf{95\%} & \textbf{80\%} & \textbf{97.71\%} & \textbf{16/20} & \textbf{17.40} \\
 & Transferred & 90\% & 70\% & 97.55\% & 15/20 & 18.86 \\
\midrule
\multirow{3}{*}{deepseek-v4-flash}
 & Base & 75\% & 70\% & 98.63\% & 16/20 & 2.99 \\
 & Direct & \textbf{85\%} & \textbf{80\%} & \textbf{98.68\%} & \textbf{17/20} & \textbf{2.84} \\
 & Transferred & 80\% & 70\% & 98.59\% & \textbf{17/20} & 3.08 \\
\bottomrule
\end{tabular}
}

\vspace{2pt}
\parbox{0.98\columnwidth}{\footnotesize
$^{*}$ Insights are learned on CRUST-Bench and transferred to SmartC2Rust-Bench. \textit{Base} uses no insights, \textit{Direct} uses target-benchmark insights, and \textit{Transferred} uses CRUST-Bench insights.}
\end{table}

Safety and idiomaticity remain stable under transfer: SafeRate changes marginally, Lint Pass is unchanged or slightly higher, and Idiom Penalty varies only modestly. Overall, \textit{Transferred} improves functional correctness without degrading code quality, while \textit{Direct} achieves the best balance.

\begin{tcolorbox}[enhanced, width=\linewidth, boxrule=0.8pt, left=2pt, right=2pt, top=2pt, bottom=2pt, drop fuzzy shadow=black,]
\textbf{Answer to RQ3.} \revtext{Transferred insights consistently match or improve the base pipeline in compilation and testing, with broadly stable safety and idiomaticity, while the full \ourmethod pipeline remains strongest due to benchmark-specific adversarial refinement.}
\end{tcolorbox}

\subsection{RQ4: Sensitivity Analysis}
\label{sec:rq4}

We study two key hyperparameters of \ourmethod on CRUST-Bench with gpt-5.4-mini: the insight retrieval size ($N$) and the maximum number of adversarial refinement rounds ($K$). To isolate their effects, we vary one parameter while fixing the other ($K{=}5$ for sweeping $N$, and $N{=}5$ for sweeping $K$).

\begin{figure}[t]
    \centering
    \subfloat[Effect of insight selection size ($N$).\label{fig:hyper-N}]{%
        \includegraphics[width=0.492\linewidth]{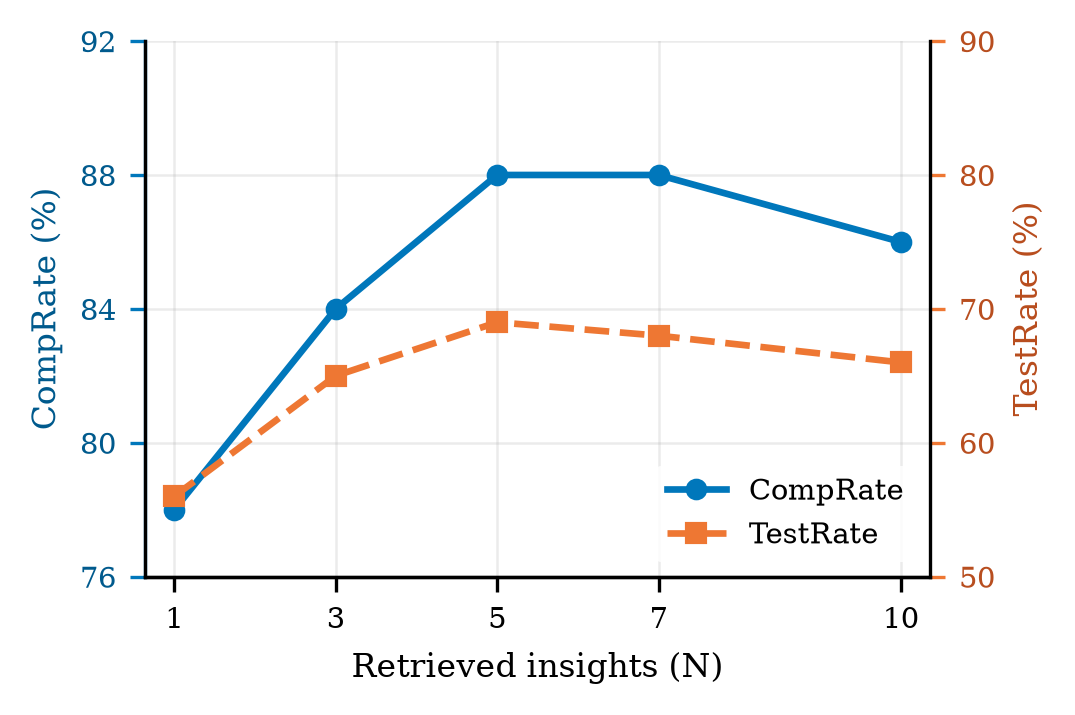}%
    }
    \hfill
    \subfloat[Effect of adversarial rounds ($K$).\label{fig:hyper-K}]{%
        \includegraphics[width=0.492\linewidth]{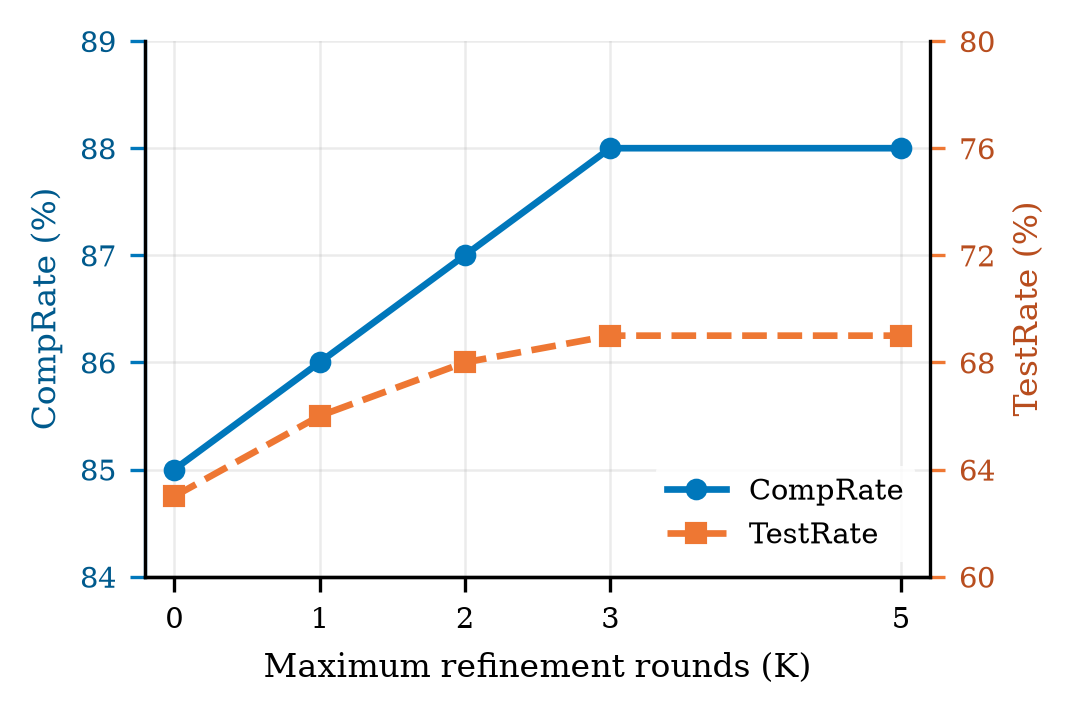}%
    }
    \caption{Hyperparameter sensitivity analysis of \ourmethod on CRUST-Bench with GPT-5.4-mini.}
    \label{fig:hyper-ablation}
\end{figure}

Figure~\ref{fig:hyper-N} shows that performance benefits from a moderate retrieval budget. With $N{=}1$, \ourmethod reaches 78\% CompRate and 56\% TestRate, indicating insufficient coverage of interacting C idioms. Increasing $N$ to 3 improves results to 84\%/65\%, and the default $N{=}5$ performs best (88\%/69\%). Larger budgets reduce TestRate (68\% at $N{=}7$, 66\% at $N{=}10$) without CompRate gains, suggesting interference from irrelevant or weakly relevant constraints.

Figure~\ref{fig:hyper-K} shows saturation in adversarial refinement. Without refinement ($K{=}0$), performance is 85\% CompRate and 63\% TestRate. One and two rounds improve results to 86\%/66\% and 87\%/68\%, while $K{\ge}3$ already matches the best observed performance (88\%/69\%). Larger $K$ mainly stabilizes the outcome.

\begin{tcolorbox}[enhanced, width=\linewidth, boxrule=0.8pt, left=2pt, right=2pt, top=2pt, bottom=2pt, drop fuzzy shadow=black,]
\textbf{Answer to RQ4.} \revtext{\ourmethod performs best under moderate retrieval and bounded adversarial learning: $N{=}5$ achieves the strongest CompRate and TestRate, while most gains are realized within two to three refinement rounds, and $K{=}5$ serves as a conservative upper bound with stable performance.}
\end{tcolorbox}

\subsection{Qualitative Insight Evolution}
\label{sec:qualitative-insight-evolution}

We inspect two representative learning traces to understand what the adversarially refined insights actually encode. 

\subsubsection{Single-Insight Adversarial Learning}
In \texttt{libvcd}, the initial pointer-buffer insight preserves buffer length and NUL termination but misses the boundary between internal storage and Rust-visible string semantics. The C code stores signals in fixed \texttt{char[N]} arrays with NUL-terminated content, while Rust conversions such as \texttt{String::from\_utf8\_lossy} may expose zero-padded regions. The challenge reveals that copying the full buffer is insufficient when downstream operations interpret it as a C string.

The refined insight \texttt{fixed-c-string-visible-length} strengthens the rule into a visibility-aware condition: fixed C arrays may retain zero-padded storage internally, but any Rust-visible representation, comparison, or lookup must stop at the first NUL unless the original program treats the buffer as raw bytes. The insight thus evolves from a structural constraint into an observation-boundary rule.

\subsubsection{Compositional Insight Adversarial Learning}
In \texttt{lib2bit}, the \texttt{twobitSequence} pipeline involves multiple cooperating insights: one preserves helper-call structure, and another enforces allocation size, decoding length, and NUL termination. Alone, they remain incomplete because one lacks the Rust-visible sequence length and the other does not specify when masking applies.

The compositional challenge exposes this interaction gap and aligns the insights into one policy: the Rust-visible length is \texttt{end - start}, the C buffer allocates one extra byte for the terminator, masking applies only to valid decoded bases, and the terminator is excluded from the logical sequence. The insights thus become a coordinated constraint set for the same abstraction.

\section{Discussion}

\subsection{Cost Analysis}
We analyze \ourmethod's cost in token usage and end-to-end runtime. On CRUST-Bench with gpt-5.4-mini, the full configuration consumes 21.43M tokens (214.3K per project), compared with 19.51M tokens (195.1K per project) for the no-insight setting. The 1.92M-token increase, or 9.8\% overhead, remains modest because most tokens are still spent on translation, compiler feedback, and test-driven repair, while the compact insight bank is injected selectively.

For runtime, \ourmethod takes 15,326 seconds in total (153.3 seconds per project), compared with 10,452 seconds (104.5 seconds per project) for SmartC2Rust, a 1.47$\times$ slowdown mainly due to trace reflection and the adversarial challenge--translate--refine loop.

These results indicate moderate token and runtime overhead, amortized as refined insights guide later translation and repair. \ourmethod is therefore most suitable for project-level migration where slightly higher latency is acceptable for improved correctness, safety, and idiomaticity, while latency-sensitive settings may rely on seeded insight banks with limited online refinement.

\subsection{Threats to Validity}
\textbf{Internal Validity}. \ourmethod validates translations using \texttt{cargo check} and \texttt{cargo test}. While these provide objective correctness signals, they do not establish full semantic equivalence with the original C implementation. Future work will complement them with fuzzing, differential testing, and stronger API specifications.

Beyond execution-based validation, compositional insight learning depends on LLM-estimated support and conflict scores in Eq.~\ref{eq:selection}, which may vary across backend models or decoding settings. We reduce this risk by using fixed schemas and prompts and by treating the scores only as a heuristic for prioritizing insight compositions before executable validation; actual insight and relation updates are accepted only when supported by compiler or test evidence.

Another threat is potential benchmark contamination during LLM pre-training. This risk is reduced because the benchmarks contain no Rust reference implementations, preventing direct memorization of target translations. However, models may still have seen the original C projects. Evaluating on private or newly released projects would further mitigate this threat.

\textbf{External Validity}. Our evaluation is limited to public user-level C projects. CRUST-Bench and SmartC2Rust-Bench cover diverse application domains, supporting generalization across a broad range of C-to-Rust translation tasks. However, they do not represent all migration scenarios, such as kernel or device-driver code, highly concurrent systems, projects with complex native dependencies, or large industrial codebases. Extending the evaluation to these settings remains future work.

\section{Related Work}

\subsection{C-to-Rust Translation}

Existing C-to-Rust translation work can be broadly grouped into rule-based and LLM-based approaches.

Rule-based methods mechanically translate C into Rust~\cite{c2rust-github,crust2019,rusty2022} while largely preserving the original program structure, often producing code that relies heavily on raw pointers and \texttt{unsafe} constructs. Subsequent work improves safety and idiomaticity through analyses and transformations for ownership~\cite{zhang2023ownership}, aliasing~\cite{emre2023aliasing}, pointer safety~\cite{emre2021safer,ling2022inrust}, API migration~\cite{concrat2023,hong2024tag,hong2024dontwrite,hong2025forcrat,genc2rust2025,chen2026inator,peng2026hayroll}, and type migration~\cite{hong2025typemigrating,xu2025typemigration}. However, these approaches rely on handcrafted analyses or transformation rules, limiting their generality.

LLM-based methods generate idiomatic Rust code without manually defined translation rules. Existing work improves translation via semantic guidance~\cite{luo2025irene,safetrans2025,yang2024vert}, retrieval~\cite{eniser2024flourine,cai2025rustmap,yuan2025ptrmapper,deptran2026}, execution-driven repair~\cite{shetty2024syzygy,zhou2025c2rusttv,rustassure2025,smartc2rust,wang2026his2trans,encrust2026}, and project-level context~\cite{orbit2026,yan2026c2rustxw,DBLP:journals/pacmse/HuZSSG26}. Despite these advances, ensuring reliable translation correctness remains challenging due to the substantial semantic and paradigm gap between C and Rust.
A line of work leverages LLMs to enhance rule-based transpilers or static analysis, for example by refining transpiler outputs~\cite{alqadiri2025translating}, injecting semantic guidance~\cite{pr22025,sactor2025}, or performing skeleton-guided repair~\cite{evoc2rust2025}.

Empirical studies~\cite{li2025userstudy,valenzuela2025fromc,tadesse2026quality,rutherford2026oxidation} and benchmarks~\cite{rustrepotrans2024,crustbench2025} further highlight persistent trade-offs among correctness, safety, and idiomaticity in C-to-Rust migration.

\ourmethod belongs to the LLM-based category. Unlike prior approaches that rely on manual rules, retrieval, or repair-driven generation, \ourmethod models reusable translation knowledge as executable insights and refines them through adversarial learning, improving the reliability of knowledge-guided project-level translation while achieving a better balance among correctness, safety, and idiomaticity.

\subsection{Experience-Enhanced AI Agents}
Experience enhancement equips LLM agents with reusable knowledge accumulated across prior tasks, enabling transfer beyond the current context. A line of work learns such experience from agent trajectories. For example, Reflexion~\cite{reflexion2023} transforms execution feedback into verbal reflections; ExpeL~\cite{expel2024} distills successful and failed trajectories into reusable lessons; AutoGuide~\cite{autoguide2024} synthesizes context-aware guidelines; and Agent Workflow Memory~\cite{agentworkflowmemory2024} extracts reusable workflows from past executions.

This paradigm has also been adopted in software engineering. AgentRR abstracts interaction traces into structured experiences for reuse across similar tasks~\cite{agentrr2025}. Agent KB aggregates heterogeneous trajectories into a knowledge base for cross-task retrieval~\cite{agentkb2025}, while SWE-Exp distills prior issue-resolution traces into actionable repair experience~\cite{sweexp2025}. These systems improve reuse by storing and retrieving past trajectories or repair patterns~\cite{DBLP:journals/corr/abs-2606-28434}, often augmented with lightweight controls such as verification~\cite{DBLP:journals/corr/abs-2606-28436}, disagreement filtering, or retrieval-based selection.

However, such experience is typically derived from stochastic agent executions, which may encode spurious decisions, omit boundary conditions, or introduce inconsistencies across tasks, limiting reliability under distribution shifts. To address this limitation, \ourmethod treats translation experience as a hypothesis to be validated: it is adversarially challenged, refined, and compositionally verified before reuse.

\section{Conclusion}
This paper presents \ourmethod, an adversarial agentic framework for C-to-Rust translation that continuously refines reusable translation insights through Translator--Challenger interaction and evidence-based updates. By strengthening insights at both individual and compositional levels, \ourmethod improves their robustness and coordination across diverse translation scenarios. 
Experiments on two project-level benchmarks show that \ourmethod consistently outperforms existing LLM-based approaches and learns insights that generalize beyond their source programs, while maintaining stable safety and idiomaticity.


\section*{Data Availability}

To support reproducibility, all related scripts and data are available at \url{https://github.com/bbzswcf/TRAIL}.

\balance
\bibliographystyle{IEEEtran}
\bibliography{IEEEabrv.bib,ref.bib}

@inproceedings{smartc2rust,
  author    = {Momoko Shiraishi and Yinzhi Cao and Takahiro Shinagawa},
  title     = {{SmartC2Rust}: Iterative, Feedback-Driven {C}-to-{Rust} Translation via Large Language Models for Safety and Equivalence},
  booktitle = {Proceedings of the ACM/IEEE 48th International Conference on Software Engineering},
  year      = {2026},
  url       = {https://arxiv.org/abs/2409.10506},
}

@inproceedings{
crustbench2025,
title={{CRUST}-Bench: A Comprehensive Benchmark for C-to-safe-Rust Transpilation},
author={Anirudh Khatry and Robert Zhang and Jia Pan and Ziteng Wang and Qiaochu Chen and Greg Durrett and Isil Dillig},
booktitle={Second Conference on Language Modeling},
year={2025},
url={https://openreview.net/forum?id=8xofWL61S9}
}

@article{alqadiri2025translating,
  title={C2 SAFERRUST: Transforming C Projects into Safer Rust with NeuroSymbolic Techniques},
  author={Nitin, Vikram and Krishna, Rahul and do Valle, Luiz Lemos and Ray, Baishakhi},
  journal={IEEE Transactions on Software Engineering},
  year={2025},
  publisher={IEEE}
}

@article{rustine2025,
  title={Translating Large-Scale C Repositories to Idiomatic Rust},
  author={Dehghan, Saman and Sun, Tianran and Wu, Tianxiang and Li, Zihan and Jabbarvand, Reyhaneh},
  journal={arXiv preprint arXiv:2511.20617},
  year={2025}
}

@inproceedings{evoc2rust2025,
  author       = {Chaofan Wang and Tingrui Yu and Beijun Shen and Jie Wang and Dong Chen and
                  Wenrui Zhang and Yuling Shi and Chen Xie and Xiaodong Gu},
  title        = {{EvoC2Rust}: {A} Skeleton-guided Framework for Project-Level {C}-to-{Rust} Translation},
  booktitle    = {IEEE/ACM International Conference on Software Engineering: Software Engineering in Practice (ICSE-SEIP)},
  year         = {2026},
  url={https://arxiv.org/abs/2508.04295}, 
}

@misc{c2rust-github,
  title        = {{C2Rust}: Migrate {C} code to {R}ust},
  year         = {2025},
  howpublished = {\url{https://github.com/immunant/c2rust}},
}

@article{emre2021safer,
  title={Translating C to safer Rust},
  author={Emre, Mehmet and Schroeder, Ryan and Dewey, Kyle and Hardekopf, Ben},
  journal={Proceedings of the ACM on Programming Languages},
  volume={5},
  number={OOPSLA},
  pages={1--29},
  year={2021},
  publisher={ACM New York, NY, USA}
}

@article{emre2023aliasing,
  title={Aliasing limits on translating C to safe Rust},
  author={Emre, Mehmet and Boyland, Peter and Parekh, Aesha and Schroeder, Ryan and Dewey, Kyle and Hardekopf, Ben},
  journal={Proceedings of the ACM on Programming Languages},
  volume={7},
  number={OOPSLA1},
  pages={551--579},
  year={2023},
  publisher={ACM New York, NY, USA}
}

@inproceedings{zhang2023ownership,
  title={Ownership guided {C} to {Rust} translation},
  author={Zhang, Hanliang and David, Cristina and Yu, Yijun and Wang, Meng},
  booktitle={International Conference on Computer Aided Verification},
  pages={459--482},
  year={2023},
  organization={Springer}
}

@inproceedings{concrat2023,
  title={Concrat: An automatic {C-to-Rust} lock {API} translator for concurrent programs},
  author={Hong, Jaemin and Ryu, Sukyoung},
  booktitle={2023 IEEE/ACM 45th International Conference on Software Engineering (ICSE)},
  pages={716--728},
  year={2023},
  organization={IEEE}
}

@inproceedings{genc2rust2025,
  title={{GenC2Rust}: Towards Generating Generic {Rust} Code from {C}},
  author={Wu, Xiafa and Demsky, Brian},
  booktitle={2025 IEEE/ACM 47th International Conference on Software Engineering (ICSE)},
  pages={90--102},
  year={2025},
  organization={IEEE}
}

@article{hong2025typemigrating,
  title={Type-migrating C-to-Rust translation using a large language model},
  author={Hong, Jaemin and Ryu, Sukyoung},
  journal={Empirical Software Engineering},
  volume={30},
  number={1},
  pages={3},
  year={2025},
  publisher={Springer}
}

@inproceedings{hong2024tag,
  title={To tag, or not to tag: Translating {C}'s unions to {Rust}'s tagged unions},
  author={Hong, Jaemin and Ryu, Sukyoung},
  booktitle={Proceedings of the 39th IEEE/ACM International Conference on Automated Software Engineering},
  pages={40--52},
  year={2024}
}

@inproceedings{hong2025forcrat,
  title={Forcrat: Automatic {I/O API} Translation from {C} to {Rust} via Origin and Capability Analysis},
  author={Hong, Jaemin and Ryu, Sukyoung},
  booktitle={2025 40th IEEE/ACM International Conference on Automated Software Engineering (ASE)},
  pages={1541--1552},
  year={2025},
  organization={IEEE}
}

@misc{pr22025,
      title={Raw Pointer Rewriting with LLMs for Translating C to Safer Rust}, 
      author={Yifei Gao and Chengpeng Wang and Pengxiang Huang and Xuwei Liu and Mingwei Zheng and Xiangyu Zhang},
      year={2026},
      eprint={2505.04852},
      archivePrefix={arXiv},
      primaryClass={cs.SE},
      url={https://arxiv.org/abs/2505.04852}, 
}

@article{eniser2024flourine,
  title={Towards translating real-world code with llms: A study of translating to rust},
  author={Eniser, Hasan Ferit and Zhang, Hanliang and David, Cristina and Wang, Meng and Christakis, Maria and Paulsen, Brandon and Dodds, Joey and Kroening, Daniel},
  journal={arXiv preprint arXiv:2405.11514},
  year={2024}
}

@inproceedings{yang2024vert,
  title={VERT: Polyglot Verified Equivalent Rust Transpilation with Large Language Models},
  author={Yang, Aidan ZH and Takashima, Yoshiki and Paulsen, Brandon and Dodds, Josiah and Kroening, Daniel},
  booktitle={2025 40th IEEE/ACM International Conference on Automated Software Engineering (ASE)},
  pages={1453--1463},
  year={2025},
  organization={IEEE}
}

@article{shetty2024syzygy,
  title={Syzygy: Dual code-test c to (safe) rust translation using llms and dynamic analysis},
  author={Shetty, Manish and Jain, Naman and Godbole, Adwait and Seshia, Sanjit A and Sen, Koushik},
  journal={arXiv preprint arXiv:2412.14234},
  year={2024}
}

@article{sactor2025,
  title={SACTOR: LLM-Driven Correct and Idiomatic C to Rust Translation with Static Analysis and FFI-Based Verification},
  author={Zhou, Tianyang and Zhang, Ziyi and Lin, Haowen and Jha, Somesh and Christodorescu, Mihai and Levchenko, Kirill and Chandrasekaran, Varun},
  journal={arXiv preprint arXiv:2503.12511},
  year={2025}
}

@inproceedings{luo2025irene,
  title={Integrating Rules and Semantics for LLM-Based C-to-Rust Translation},
  author={Luo, Feng and Ji, Kexing and Gao, Cuiyun and Gao, Shuzheng and Feng, Jia and Liu, Kui and Xia, Xin and Lyu, Michael R},
  booktitle={2025 IEEE International Conference on Software Maintenance and Evolution (ICSME)},
  pages={685--696},
  year={2025},
  organization={IEEE}
}

@inproceedings{cai2025rustmap,
  title={RustMap: Towards Project-Scale C-to-Rust Migration via Program Analysis and LLM},
  author={Cai, Xuemeng and Liu, Jiakun and Huang, Xiping and Yu, Yijun and Wu, Haitao and Li, Chunmiao and Wang, Bo and Yusuf, Imam Nur Bani and Jiang, Lingxiao},
  booktitle={International Conference on Engineering of Complex Computer Systems},
  pages={283--302},
  year={2025}
}

@inproceedings{safetrans2025,
  title={SafeTrans: LLM-assisted Transpilation from C to Rust},
  author={Farrukh, Muhammad and Coskun, Baris and Palit, Tapti and Polychronakis, Michalis},
  booktitle={Proceedings of the 1st Workshop on Code Translation, Transformation, and Modernization},
  pages={30--37},
  year={2026}
}

@misc{yuan2025ptrmapper,
      title={Project-Level C-to-Rust Translation via Pointer Knowledge Graphs}, 
      author={Zhiqiang Yuan and Wenjun Mao and Zhuo Chen and Xiyue Shang and Chong Wang and Yiling Lou and Xin Peng},
      year={2026},
      eprint={2510.10956},
      archivePrefix={arXiv},
      primaryClass={cs.SE},
      url={https://arxiv.org/abs/2510.10956}, 
}

@inproceedings{zhou2025c2rusttv,
  title={C2rusttv: An llm-based framework for c to rust translation and validation},
  author={Zhou, Han and Luo, Yu and Zhang, Mengtao and Xu, Dianxiang},
  booktitle={2025 IEEE 49th Annual Computers, Software, and Applications Conference (COMPSAC)},
  pages={1254--1259},
  year={2025},
  organization={IEEE}
}

@inproceedings{valenzuela2025fromc,
  title={From C to Rust: Evaluating LLM Capabilities in Transpilation Through Compilation Errors},
  author={Valenzuela, Andrea and Gonzalez-Mallo, Marta and Gutierrez, Cristian and Garcia-Gasulla, Dario and Kestor, Gokcen and Royuela, Sara},
  booktitle={International Conference on High Performance Computing},
  pages={311--324},
  year={2025},
  organization={Springer}
}

@inproceedings{xu2025typemigration,
  title={Optimizing type migration for llm-based c-to-rust translation: A data flow graph approach},
  author={Xu, Qingxiao and Huang, Jeff},
  booktitle={Proceedings of the 14th ACM SIGPLAN International Workshop on the State Of the Art in Program Analysis},
  pages={8--14},
  year={2025}
}

@article{tadesse2026quality,
  title={Code Quality Analysis of Translations from C to Rust},
  author={Tadesse, Biruk and Nitin, Vikram and Salah, Mazin and Ray, Baishakhi and d'Amorim, Marcelo and Assun{\c{c}}{\~a}o, Wesley},
  journal={arXiv preprint arXiv:2602.00840},
  year={2026}
}

@article{wang2026his2trans,
  title={Build-Aware Incremental C-to-Rust Migration via Skeleton-First Translation and Historical Knowledge Reuse},
  author={Wang, Shengbo and Liu, Mingwei and Ou, Guangsheng and Chen, Yuwen and Li, Zike and Wang, Yanlin and Zheng, Zibin},
  journal={arXiv preprint arXiv:2603.02617},
  year={2026}
}

@article{yan2026c2rustxw,
  title={C2RustXW: Program-Structure-Aware C-to-Rust Translation via Program Analysis and LLM},
  author={Yan, Yanyan and Feng, Yang and Liu, Jiangshan and Liu, Di and Liu, Zixi and Teng, Hao and Xu, Baowen},
  journal={arXiv preprint arXiv:2603.28686},
  year={2026}
}

@article{encrust2026,
  title={ENCRUST: Encapsulated Substitution and Agentic Refinement on a Live Scaffold for Safe C-to-Rust Translation},
  author={Sim, Hohyun and Cho, Hyeonjoong and Shokri, Ali and Fu, Zhoulai and Ravindran, Binoy},
  journal={arXiv preprint arXiv:2604.04527},
  year={2026}
}

@article{chen2026inator,
  title={\&inator: Correct, Precise C-to-Rust Interface Translation},
  author={Chen, Victor and Coughlin, Ayden and Bond, Michael D},
  journal={Proceedings of the ACM on Programming Languages},
  volume={10},
  number={PLDI},
  pages={580--603},
  year={2026},
  publisher={ACM New York, NY, USA}
}

@article{peng2026hayroll,
  title={Hayroll: A Modular Wrapper for Translating C Macros and Conditional Compilation to Rust},
  author={Peng, Haoran and Kasikci, Baris and Bernstein, Gilbert Louis and Ernst, Michael D},
  journal={Proceedings of the ACM on Programming Languages},
  volume={10},
  number={PLDI},
  pages={730--753},
  year={2026},
  publisher={ACM New York, NY, USA}
}

@inproceedings{rutherford2026oxidation,
  title={An Empirical Study of C to Rust Translation using Local Large-Language Models},
  author={Rutherford, Nathan and O’Keeffe, Dan},
  booktitle={The Third International Workshop on Large Language Models for Code},
  year={2026}
}

@incollection{crust2019,
  title={Crust: AC/C++ to Rust transpiler using a “nano-parser methodology” to avoid C/C++ safety issues in legacy code},
  author={Shetty, Nishanth and Saldanha, Nikhil and Thippeswamy, MN},
  booktitle={Emerging Research in Computing, Information, Communication and Applications: ERCICA 2018, Volume 1},
  pages={241--250},
  year={2019},
  publisher={Springer}
}

@inproceedings{rusty2022,
  title={RUSTY: Effective C to Rust conversion via unstructured control specialization},
  author={Han, Xiangjun and Hua, Baojian and Wang, Yang and Zhang, Ziyao},
  booktitle={2022 IEEE 22nd International Conference on Software Quality, Reliability, and Security Companion (QRS-C)},
  pages={760--761},
  year={2022},
  organization={IEEE}
}

@inproceedings{rustrepotrans2024,
  title={RustRepoTrans: Repository-level Context Code Translation Benchmark Targeting Rust},
  author={Ou, Guangsheng and Liu, Mingwei and Chen, Yuxuan and Wang, Yanlin and Peng, Xin and Zheng, Zibin},
  booktitle={2025 40th IEEE/ACM International Conference on Automated Software Engineering (ASE)},
  pages={610--622},
  year={2025},
  organization={IEEE}
}

@article{li2025userstudy,
  title={Translating c to rust: Lessons from a user study},
  author={Li, Ruishi and Wang, Bo and Li, Tianyu and Saxena, Prateek and Kundu, Ashish},
  journal={arXiv preprint arXiv:2411.14174},
  year={2024}
}

@inproceedings{rustassure2025,
  title={RustAssure: Differential Symbolic Testing for LLM-Transpiled C-to-Rust Code},
  author={Bai, Yubo and Palit, Tapti},
  booktitle={2025 40th IEEE/ACM International Conference on Automated Software Engineering (ASE)},
  pages={534--546},
  year={2025},
  organization={IEEE}
}

@article{deptran2026,
  title={Dependency-Guided Repository-Level C-to-Rust Translation with Reinforcement Alignment},
  author={Feng, Jia and Gan, Wenjie and Gao, Cuiyun and Wang, Chaozheng and Luo, Feng and Xia, Xin and Li, Ge and Liu, Kui},
  journal={arXiv preprint arXiv:2604.02852},
  year={2026}
}

@misc{orbit2026,
      title={ORBIT: Guided Agentic Orchestration for Autonomous C-to-Rust Transpilation}, 
      author={Muhammad Farrukh and Baris Coskun and Tapti Palit and Michalis Polychronakis},
      year={2026},
      eprint={2604.12048},
      archivePrefix={arXiv},
      primaryClass={cs.SE},
      url={https://arxiv.org/abs/2604.12048}, 
}

@inproceedings{ling2022inrust,
  title={In rust we trust: a transpiler from unsafe c to safer rust},
  author={Ling, Michael and Yu, Yijun and Wu, Haitao and Wang, Yuan and Cordy, James R and Hassan, Ahmed E},
  booktitle={Proceedings of the ACM/IEEE 44th international conference on software engineering: companion proceedings},
  pages={354--355},
  year={2022}
}

@article{hong2024dontwrite,
  title={Don’t write, but return: Replacing output parameters with algebraic data types in c-to-rust translation},
  author={Hong, Jaemin and Ryu, Sukyoung},
  journal={Proceedings of the ACM on Programming Languages},
  volume={8},
  number={PLDI},
  pages={716--740},
  year={2024},
  publisher={ACM New York, NY, USA}
}

@article{reflexion2023,
  title={Reflexion: Language agents with verbal reinforcement learning},
  author={Shinn, Noah and Cassano, Federico and Gopinath, Ashwin and Narasimhan, Karthik and Yao, Shunyu},
  journal={Advances in neural information processing systems},
  volume={36},
  pages={8634--8652},
  year={2023}
}

@inproceedings{expel2024,
  title={Expel: Llm agents are experiential learners},
  author={Zhao, Andrew and Huang, Daniel and Xu, Quentin and Lin, Matthieu and Liu, Yong-Jin and Huang, Gao},
  booktitle={Proceedings of the AAAI Conference on Artificial Intelligence},
  volume={38},
  number={17},
  pages={19632--19642},
  year={2024}
}

@article{sweexp2025,
  title={Swe-exp: Experience-driven software issue resolution},
  author={Chen, Silin and Lin, Shaoxin and Shi, Yuling and Lian, Heng and Gu, Xiaodong and Yun, Longfei and Chen, Dong and Cao, Lin and Liu, Jiyang and Xia, Nu and others},
  journal={arXiv preprint arXiv:2507.23361},
  year={2025}
}

@article{agentworkflowmemory2024,
  title={Agent workflow memory},
  author={Wang, Zora Zhiruo and Mao, Jiayuan and Fried, Daniel and Neubig, Graham},
  journal={arXiv preprint arXiv:2409.07429},
  year={2024}
}

@article{autoguide2024,
  title={Autoguide: Automated generation and selection of context-aware guidelines for large language model agents},
  author={Fu, Yao and Kim, Dong-Ki and Kim, Jaekyeom and Sohn, Sungryull and Logeswaran, Lajanugen and Bae, Kyunghoon and Lee, Honglak},
  journal={Advances in Neural Information Processing Systems},
  volume={37},
  pages={119919--119948},
  year={2024}
}

@article{agentrr2025,
  title={Get experience from practice: Llm agents with record \& replay},
  author={Feng, Erhu and Zhou, Wenbo and Liu, Zibin and Chen, Le and Dong, Yunpeng and Zhang, Cheng and Zhao, Yisheng and Du, Dong and Hua, Zhichao and Xia, Yubin and others},
  journal={arXiv preprint arXiv:2505.17716},
  year={2025}
}

@article{agentkb2025,
  title={Agent kb: Leveraging cross-domain experience for agentic problem solving},
  author={Tang, Xiangru and Qin, Tianrui and Peng, Tianhao and Zhou, Ziyang and Shao, Daniel and Du, Tingting and Wei, Xinming and Xia, Peng and Wu, Fang and Zhu, He and others},
  journal={arXiv preprint arXiv:2507.06229},
  year={2025}
}

@article{zeng2025pruning,
  title={Pruning the Unsurprising: Efficient LLM Reasoning via First-Token Surprisal},
  author={Zeng, Wenhao and Wang, Yaoning and Hu, Chao and Shi, Yuling and Wan, Chengcheng and Zhang, Hongyu and Gu, Xiaodong},
  journal={arXiv preprint arXiv:2508.05988},
  year={2025}
}

@inproceedings{DBLP:conf/acl/ZengZSHCSG26,
  author       = {Wenhao Zeng and
                  Xuteng Zhang and
                  Yuling Shi and
                  Chao Hu and
                  Yuting Chen and
                  Beijun Shen and
                  Xiaodong Gu},
  editor       = {Maria Liakata and
                  Viviane P. Moreira and
                  Jiajun Zhang and
                  David Jurgens},
  title        = {GlimpRouter: Efficient Collaborative Inference by Glimpsing One Token
                  of Thoughts},
  booktitle    = {Findings of the Association for Computational Linguistics, {ACL} 2026,
                  San Diego, California, United States, July 2-7, 2026},
  pages        = {17850--17864},
  publisher    = {Association for Computational Linguistics},
  year         = {2026},
  url          = {https://doi.org/10.18653/v1/2026.findings-acl.885},
  doi          = {10.18653/V1/2026.FINDINGS-ACL.885},
  bibsource    = {dblp computer science bibliography, https://dblp.org}
}

@article{zhang2026paratempo,
  title={ParaTempo: Efficient Parallel Reasoning via Temporal Confidence},
  author={Zhang, Xuteng and Zeng, Wenhao and Gu, Xiaodong and Hu, Chao and Lin, Haotian and Shi, Yuling and Wang, Min and Shen, Beijun},
  journal={arXiv preprint arXiv:2608.16425},
  year={2026}
}

@article{DBLP:journals/corr/abs-2606-28436,
  author       = {Wenhao Zeng and
                  Yuling Shi and
                  Xiaodong Gu and
                  Chao Hu and
                  Chaofan Wang and
                  Yuhao Cui and
                  Hongting Zhou and
                  Mengnan Qi and
                  Jianqiao Wangni and
                  Zhaojian Yu and
                  Shuzheng Gao and
                  Kai Cai and
                  Shilin He},
  title        = {Dockerless: Environment-Free Program Verifier for Coding Agents},
  journal      = {CoRR},
  volume       = {abs/2606.28436},
  year         = {2026},
  url          = {https://doi.org/10.48550/arXiv.2606.28436},
  doi          = {10.48550/ARXIV.2606.28436},
  eprinttype   = {arXiv},
  eprint       = {2606.28436},
  bibsource    = {dblp computer science bibliography, https://dblp.org}
}

@article{DBLP:journals/pacmse/HuZSSG26,
  author       = {Chao Hu and
                  Wenhao Zeng and
                  Yuling Shi and
                  Beijun Shen and
                  Xiaodong Gu},
  title        = {In Line with Context: Repository-Level Code Generation via Context
                  Inlining},
  journal      = {Proc. {ACM} Softw. Eng.},
  volume       = {3},
  number       = {{FSE}},
  pages        = {1469--1491},
  year         = {2026},
  url          = {https://doi.org/10.1145/3797094},
  doi          = {10.1145/3797094},
  bibsource    = {dblp computer science bibliography, https://dblp.org}
}

@article{DBLP:journals/corr/abs-2606-28434,
  author       = {Shuzheng Gao and
                  Wenhao Zeng and
                  Zhaojian Yu and
                  Jianqiao Wangni and
                  Chaozheng Wang and
                  Kai Cai and
                  Shilin He and
                  Michael R. Lyu},
  title        = {SWE-MeM: Learning Adaptive Memory Management for Long-Horizon Coding
                  Agents},
  journal      = {CoRR},
  volume       = {abs/2606.28434},
  year         = {2026},
  url          = {https://doi.org/10.48550/arXiv.2606.28434},
  doi          = {10.48550/ARXIV.2606.28434},
  eprinttype   = {arXiv},
  eprint       = {2606.28434},
  bibsource    = {dblp computer science bibliography, https://dblp.org}
}

\end{document}